\documentclass[prd,superscriptaddress,amsfonts,amssymb,amsmath,showpacs,twocolumn]{revtex4-2}
\usepackage{bm}
\usepackage{amsfonts}
\usepackage{latexsym}
\usepackage{graphicx}
\usepackage{amsmath}
\usepackage{palatino}
\usepackage{mathpazo}
\usepackage{textcomp}
\usepackage{float}
\usepackage{booktabs}
\usepackage{dcolumn}
\usepackage{booktabs}
\usepackage{multirow}
\usepackage{hyperref}
\hypersetup{colorlinks,citecolor=blue}
\usepackage{amsmath}
\usepackage{xcolor}
\usepackage{orcidlink}
\usepackage[caption=false]{subfig}
\usepackage{commath}
\def\jnl@style{\it}
\def\aaref@jnl#1{{\jnl@style#1}}

\def\aaref@jnl#1{{\jnl@style#1}}

\def\aj{\aaref@jnl{AJ}}                   
\def\apj{\aaref@jnl{ApJ}}                 
\def\apjl{\aaref@jnl{ApJ}}                
\def\apjs{\aaref@jnl{ApJS}}               
\def\apss{\aaref@jnl{Ap\&SS}}             
\def\aap{\aaref@jnl{A\&A}}                
\def\aapr{\aaref@jnl{A\&A~Rev.}}          
\def\aaps{\aaref@jnl{A\&AS}}              
\def\mnras{\aaref@jnl{Mon.~Not.~Roy.~Astron.~Soc.}}             
\def\prd{\aaref@jnl{Phys.~Rev.~D}}        
\def\prc{\aaref@jnl{Phys.~Rev.~C}}  
\def\prl{\aaref@jnl{Phys.~Rev.~Lett.}}    
\def\qjras{\aaref@jnl{QJRAS}}             
\def\skytel{\aaref@jnl{S\&T}}             
\def\ssr{\aaref@jnl{Space~Sci.~Rev.}}     
\def\zap{\aaref@jnl{ZAp}}                 
\def\nat{\aaref@jnl{Nature}}              
\def\aplett{\aaref@jnl{Astrophys.~Lett.}} 
\def\apspr{\aaref@jnl{Astrophys.~Space~Phys.~Res.}} 
\def\physrep{\aaref@jnl{Phys.~Rep.}}      
\def\physscr{\aaref@jnl{Phys.~Scr}}       
\def\commat{\aaref@jnl{Comm.~Math.~Phys.}}              
\def\science{\aaref@jnl{Science}}               
\def\cqg{\aaref@jnl{Classical Quant.~Grav.}}            
\def\jpcs{\aaref@jnl{JPCS}}                                     
\def\ijmpd{\aaref@jnl{Int.~J.~Mod.~Phys.~D}}                    
\def\grg{\aaref@jnl{Gen.~Relat.~Gravit.}}               
\def\rpp{\aaref@jnl{Rep.~Prog.~Phys.}}          
\def\npa{\aaref@jnl{Nucl.~Phys.~A}}        
\def\lrr{\aaref@jnl{Living Rev.~Rel.}}                   
\def\jcap{\aaref@jnl{J.~Cosmology Astropart.~Phys.}}    
\def\rmp{\aaref@jnl{Rev.~Mod.~Phys.}}   
\def\epjc{\aaref@jnl{Eur.~Phys.~J.~C}}

\allowdisplaybreaks[1]
\renewcommand{\arraystretch}{1.1}
\begin{document}

\color{black}       

\title{Impact of dark energy on the equation of state in light of the latest cosmological data}

\author{N. Myrzakulov\orcidlink{0000-0001-8691-9939}}
\email[Email: ]{nmyrzakulov@gmail.com}
\affiliation{L. N. Gumilyov Eurasian National University, Astana 010008,
Kazakhstan.}
\affiliation{Ratbay Myrzakulov Eurasian International Centre for Theoretical
Physics, Astana 010009, Kazakhstan.}

\author{M. Koussour\orcidlink{0000-0002-4188-0572}}
\email[Email: ]{pr.mouhssine@gmail.com}
\affiliation{Quantum Physics and Magnetism Team, LPMC, Faculty of Science Ben
M'sik,\\
Casablanca Hassan II University,
Morocco.}

\author{Alnadhief H. A. Alfedeel\orcidlink{0000-0002-8036-268X}}%
\email[Email: ]{aaalnadhief@imamu.edu.sa}
\affiliation{Department of Mathematics and Statistics, Imam Mohammad Ibn Saud Islamic University (IMSIU),\\
Riyadh 13318, Saudi Arabia.}
\affiliation{Department of Physics, Faculty of Science, University of Khartoum, P.O. Box 321, Khartoum 11115, Sudan.}
\affiliation{Centre for Space Research, North-West University, Potchefstroom 2520, South Africa.}

\author{E. I. Hassan\orcidlink{0000-0000-0000-0000}}%
\email[Email:]{eiabdalla@imamu.edu.sa}
\affiliation{Department of Mathematics and Statistics, Imam Mohammad Ibn Saud Islamic University (IMSIU),\\
Riyadh 13318, Saudi Arabia.}


\date{\today}

\begin{abstract}
We reconstruct the effective Equation of State (EoS) within the framework of General Relativity (GR) theory in a homogeneous and isotropic FLRW Universe, which is assumed to be composed of matter and Dark Energy (DE). Our analysis employs a dataset consisting of 31 Cosmic Chronometer (CC) data points, six data points of Baryon Acoustic Oscillations (BAO), and 1048 Type Ia Supernovae (SN) from the Pantheon sample, and we determine the best-fit values of the model parameters through Markov Chain Monte Carlo (MCMC) simulation. We then use these parameter values to calculate various cosmological parameters, such as the DE EoS parameter, the energy density, the deceleration parameter, the state finder parameters, and the $Om(z)$ diagnostic. All the analyzed cosmological parameters show behavior consistent with the accelerated Universe scenario.
\end{abstract}

\maketitle

\section{Introduction}
\label{sec1}

Observational data suggest that the Universe underwent two distinct periods of cosmic acceleration. The first is called inflation, and it started shortly following the Big Bang. The second, known as a protracted cosmic acceleration, began approximately 6 billion years after the Big Bang and continues to this day. Observations of Type Ia supernovae (SN) in the late 20th century revealed that the expansion of the present Universe is accelerating \cite{Riess, Perlmutter}, which contradicts the General Theory of Relativity (GR). Additional data from observations of the Baryon Acoustic Oscillations (BAO) \cite{D.J., W.J.}, the Cosmic Microwave Background (CMB) \cite{R.R., Z.Y.}, the Large Scale Structure (LSS) \cite{T.Koivisto, S.F.}, the recent Planck collaboration \cite{Planck2020}, and other sources have since confirmed this late-time cosmic acceleration. Perlmutter et al. were awarded the 2011 \textit{Nobel Prize in Physics} for their contributions to this discovery \cite{Riess, Perlmutter}. While the cause of cosmic acceleration is still unknown, most cosmologists believe that it is due to "Dark Energy" (DE), which accounts for approximately 70\% of the Universe's current energy budget and is characterized by negative pressure. Measurements of the large-scale clustering \cite{LSC1,LSC2}, cosmic age \cite{CA}, weak lensing \cite{WL}, and gamma-ray bursts \cite{GRB1,GRB2} have further constrained the nature of DE. The DE Equation of State (EoS) parameter $\omega$, which represents the ratio of pressure $p$ to energy density $\rho$, must be less than $-1/3$ for late-time acceleration to occur.

In light of recent observational evidence, various models of DE have been proposed, with the simplest being the $\Lambda$CDM ($\Lambda$ Cold Dark Matter) model. In this model, the DE is represented by the cosmological constant $\Lambda$, whose energy density remains constant over time and has an EoS parameter of $\omega_{\Lambda}=-1$, i.e. $p_{\Lambda}=-\rho_{\Lambda}$. However, this model suffers from the fine-tuning problem and the cosmological coincidence problem \cite{dalal/2001, weinberg/1989}. To overcome these problems, it is necessary to consider alternative theories that provide an explanation for the origin and properties of DE. These theoretical issues have motivated cosmologists to explore other unidentified components that can be responsible for the Universe's late-time accelerated expansion. Scalar field models have gained considerable popularity as potential candidates for DE due to their simple and dynamic nature. In order to explain the Universe's accelerated expansion in recent times, various dynamical DE models have been studied over the past decade. These include quintessence models where DE is linked to a time-dependent canonical scalar field, k-essence, phantom models, quintom models, tachyon models, chaplygin gas models, and others \cite{Qu3,Qu4,Phantom1,Phantom2,T.Chiba,C.Arm.,Kamenshchik,M.C.,A.Y.,tachyon}. Modified gravity theories offer an alternative explanation for the accelerated expansion of the Universe and have gained popularity in recent years. Rather than invoking a physical DE component, these theories modify the Einstein-Hilbert action of gravity by introducing functions of specific combinations of curvature invariants, such as the Riemann tensor, the Ricci tensor, and the Weyl tensor. Examples of such modifications include $f(R)$ gravity (where $R$ denotes the curvature scalar) \cite{fR1,fR2}, $f(T)$ gravity (where $T$ denotes the torsion scalar) \cite{fT1,fT2}, $f(Q)$ gravity (where $Q$ denotes the non-metricity scalar) \cite{fQ1,fQ2,fQ3,fQ4,fQ5,fQ6,fQ7}, and others. 

Despite extensive research, the cause and nature of DE remain largely unknown. A popular method for modeling late-time cosmic acceleration is known as "reconstruction", which involves building a model based on direct observational data. This approach is essentially the reverse of the traditional method for finding a suitable cosmological model \cite{Mukherjee}. The idea of using scalar field potential for reconstruction was introduced by Ellis and Madsen \cite{Ellis}. Two types of reconstruction exist, parametric and non-parametric. Parametric reconstruction (or DE parametrization method) involves estimating model parameters from observational data, with the goal of assuming a particular evolution scenario and identifying the matter sector or exotic component responsible for the acceleration. Examples of studies using parametric reconstruction include those by Starobinsky \cite{Starobinsky}, Huterer and Turner \cite{Huterer1,Huterer2}, and Saini et al. \cite{Saini}. More recent studies have explored different parametrizations of DE EoS. The most commonly used parametrization is the Chevallier-Polarski-Linder (CPL) parametrization i.e. $\omega=\omega_{0}+\omega_{1}\frac{z}{1+z}$, which is based on a simple Taylor expansion of the EoS in terms of the scale factor \cite{CPL1,CPL2}. Other popular parametrizations include the Jassal-Bagla-Padmanabhan (JBP) parametrization i.e. $\omega=\omega_{0}+\omega_{1}\frac{z}{(1+z)^2}$, which allows for a transition from a DE-dominated Universe to a matter-dominated Universe \cite{JBP}, and the Ma-Zhang (MZ) parametrization, which is based on a logarithmic and oscillating form of the EoS, i.e. $\omega=\omega_{0}+\omega_{1}(\frac{\ln(2+z)}{1+z}-\ln2)$ and $\omega=\omega_{0}+\omega_{1}(\frac{\sin(1+z)}{1+z}-\sin(2))$, respectively \cite{MZ}. Recently, Mukherjee \cite{Mukherjee} presents a parametric form of the EoS and performs a reconstruction analysis to constrain the model parameters using cosmological observational data. The aim is to reconstruct the effective EoS and investigate its behavior over cosmic time. The study finds that a time-dependent EoS is favored by the observational data, indicating that the DE component is not a cosmological constant but rather a dynamical scalar field. The paper by Mamon \cite{Mamon} presents a reconstruction of the interaction rate in the holographic DE model, where the Hubble horizon is considered as the infrared cut-off. The author uses observational data from the Hubble parameter measurements and the BAO data to constrain the model parameters and reconstruct the interaction rate. The results suggest that the interaction rate is non-zero, and the holographic DE model with a non-zero interaction rate is consistent with the observational data. These parametrizations have been extensively tested using various observational data sets, and have been shown to provide a good fit to the data. However, it is important to keep in mind that these parametrizations are not based on any specific theoretical framework, and may not capture the true behavior of DE. In contrast, non-parametric reconstruction is a statistical analysis method for observational data that does not rely on any prior assumptions about the parametric form of any cosmological parameter. This approach involves a rigorous examination of the data to extract information without imposing any preconceived models or assumptions. The primary goal of non-parametric reconstruction is to directly identify the nature of the Universe's history from observational data, as demonstrated in studies by Holsclaw et al. \cite{Holsclaw1,Holsclaw2}, Crittenden et al. \cite{Crittenden}, and Nair et al \cite{Nair}.

In this study, a method for reconstructing the effective (or total) EoS using parametric techniques has been proposed. The chosen functional form of the effective EoS parameter is designed to approach zero at high redshifts, indicating a matter-dominated Universe. The value of the effective EoS parameter at present depends on the model parameters i.e. $\alpha$ and $\beta$, which have been constrained using observational data. To obtain the posterior probability distribution of the model parameters, we have employed the $\chi^2$ minimization technique, equivalent to maximum-likelihood analysis, and the Markov Chain Monte Carlo (MCMC) approach. Several observational datasets, including the 31 Cosmic Chronometer
(CC) data points, six BAO points, and 1048 SN from the Pantheon sample, were used to obtain the constraints on the model parameters.
 
The organization of the paper is as follows: Sec. \ref{sec2} presents the mathematical formulation of the effective EoS reconstruction. The best-fit values of the model parameters are determined in Sec. \ref{sec3} using the CC, BAO, and CC+BAO+SN datasets. The statistical analysis results and the behavior of the cosmological parameters for the model parameters constrained by observational datasets are examined in Sec \ref{sec4}. In Sec. \ref{sec5}, in addition to the deceleration parameter, statefinder and $Om(z)$ diagnostics are applied to differentiate the present $\omega_{eff}$ cosmological model from other DE models. Finally, the findings are discussed and concluded in Sec. \ref{sec5}.

\section{Reconstruction of the Cosmological Model}
\label{sec2}

The Friedmann equations, based on the FLRW (Friedmann-Lemaître-Robertson-Walker) metric, provides the basic mathematical framework for modern cosmology. In the case of a spatially flat Universe, the FLRW metric takes the form:
\begin{equation}
ds^{2}=-dt^{2}+a^{2}(t)\left[ dr^{2}+r^{2}\left( d\theta ^{2}+\sin
^{2}\theta d\psi ^{2}\right) \right] .
\label{FLRW}
\end{equation}
where $a(t)$ is the scale factor, which describes the evolution of the distances between any two points in the Universe over time, $r$, $\theta$, and $\psi$ denote the co-moving radial and angular coordinates, respectively. The Hubble parameter $H$ is defined as $H = \frac{\dot{a}}{a}$, where an over-dot denotes a derivative with respect to cosmic time $t$. The Ricci scalar for the FLRW metric shown above can be calculated using the formula $R=-6(\dot{H}+2H^{2})$. 

The energy-momentum tensor for a perfect fluid is given by
\begin{equation}
T_{\mu \nu }=(\rho +p)u_{\mu
}u_{\nu }+pg_{\mu \nu },
\label{EMT}
\end{equation}
where $\rho$ is the total energy density, $p$ is the pressure, $u_{\mu}$ is the 4-velocity of the fluid, and $g_{\mu\nu}$ is the metric tensor.

Incorporating the FLRW metric into Einstein's field equations yields the Friedmann equations, which describe the evolution of the Universe over time. For a spatially flat Universe, the Friedmann equations  with a perfect fluid take the form:
\begin{equation}
3H^2 = 8\pi G\rho,
\label{F1}
\end{equation}
\begin{equation}
2\dot{H} + 3H^2 = -8\pi Gp,
\label{F2}
\end{equation}
where $G$ is the gravitational constant. Throughout this discussion, it will be assumed that natural units are used, where the gravitational constant times $8\pi$, $G$, is set to unity ($8\pi G = 1$). The first equation, known as the energy density equation, relates the energy density of matter and radiation to the expansion rate of the Universe. The second equation, known as the acceleration equation, relates the acceleration of the expansion rate to the pressure of matter and radiation.

To describe the properties of DE, a new component of the Universe responsible for the observed acceleration, the concept of an effective equation of state (EoS) parameter $\omega_{eff}$ is introduced. This parameter is defined as the relationship between pressure and energy density, i.e.,
\begin{equation}
    \omega_{eff}=\frac{p}{\rho}. 
\end{equation}

It should be noted that the energy density $\rho$ and pressure $p$, defined in the Friedmann equations, account for the density and pressure of all types of matter present in the Universe such as matter ($\rho_{m}$ and $p_{m}=0$) and DE ($\rho_{DE}$ and $p_{DE}$).

From Eqs. (\ref{F1}) and (\ref{F2}), the effective EoS parameter can be expressed using the Hubble parameter $H$ as,
\begin{equation}
    \omega_{eff}=-1-\frac{2}{3}\left( \frac{\dot{H}}{H^2} \right).
    \label{omegaH}
\end{equation}

To make it easier to compare the theoretical results with observations, we introduce a new independent variable, the redshift $z$, instead of the usual time variable $t$. The redshift is defined as,
\begin{equation}
    z=\frac{a_{0}}{a(t)}-1,
\end{equation}
where we can simplify our analysis by normalizing the scale factor such that its present-day value is equal to one, i.e., $a(0)=1$. Thus, we can express the derivatives with respect to cosmic time in terms of derivatives with respect to redshift using the relation:
\begin{equation}
    \frac{d}{dt}=-\left(
1+z\right) H\left( z\right) \frac{d }{dz}.
\end{equation}

The derivative with respect to cosmic time of the Hubble parameter $H$ can be written as a function of the redshift as,
\begin{equation}
\overset{.}{H}=-\left(
1+z\right) H\left( z\right) \frac{dH\left( z\right) }{dz}.
\label{t_z}
\end{equation}

The conservation equations governing the dynamics of the DE and matter field can be expressed as,
\begin{equation}
\overset{.}{\rho }_{m}+3\rho _{m}H=0,  \label{cm}
\end{equation}%
\begin{equation}
\overset{.}{\rho }_{DE}+3\left( \rho _{DE}+p_{DE}\right) H=0.
\label{cf}
\end{equation}

The energy density of matter can be obtained by solving Eq. (\ref{cm}). The solution is given by
\begin{equation}
\rho _{m}=\rho _{m0}\left( 1+z\right) ^{3},
\label{rhom}
\end{equation}%
where $\rho_{m0}$ is the constant of integration that denotes the energy density of matter in the present-day. 

To close the system of equations represented by Eqs. (\ref{F1}) and (\ref{F2}), an ansatz or assumption is needed. In this particular work, the assumption is made that the effective EoS parameter can be expressed as an exponential function of the redshift $z$. This parametric form is given by:
\begin{equation}
    \omega_{eff}(z)=-1+\frac{1}{1+\frac{\alpha }{\beta +z}e^{-z}},
    \label{exp}
\end{equation}
where $\alpha$ and $\beta$ are model parameters to be determined by fitting the model to data. The exponential form for the effective EoS parameter can potentially capture a wide range of behaviors for the EoS parameter, including possible deviations from a smooth evolution, and it allows for a flexible description of the DE component. It is worth noting that the proposed form of the effective EoS contains an exponential function that introduces a complex relationship with the redshift. This aspect should be taken into consideration because it may be challenging to constrain the model parameters with the existing cosmological observations. This is due to the fact that it adds more degrees of freedom to the model, and the accuracy of the results may depend on the dataset used for the fitting. For this purpose, we approximate the exponential function using the Taylor series expansion as $e^{z}\sim1+z$.  In this case, the effective EoS parameter as presented in Eq. (\ref{exp}), takes the following form:
\begin{equation}
    \omega_{eff}(z)=-\frac{\alpha }{\alpha +(1+z) (\beta +z)}.
    \label{eff}
\end{equation}

The motivation for the effective EoS parameter $\omega_{eff}$, given by Eq. (\ref{eff}) is rooted in observations of the LSS of the Universe and the current understanding of the dominant component of the Universe. At high redshift, the dominant component of the Universe is believed to be matter, which is a pressureless component, and thus the effective EoS parameter is effectively zero. However, in the present era, the Universe is dominated by a mysterious component known as DE, which is believed to be responsible for the observed accelerating expansion of the Universe. The effective EoS parameter for DE is known to be negative and less than $-1/3$, and can be described by a time-dependent EoS. The chosen functional form of the effective EoS parameter is an exponential function of the redshift $z$, with parameters $\alpha$ and $\beta$. This functional form was chosen because it is able to capture the two phases of the evolution of the Universe, where the Universe is dominated by matter at high redshifts i.e. at $z\to\infty$, $\omega_{eff}=0$, and by DE at low redshifts i.e. at $z=0$ (present) and $z\to-1$, $\omega_{eff}=-\frac{\alpha }{\alpha +\beta }$ and $\omega_{eff}=-1$, respectively. Thus, at
the present epoch, the value of $\omega_{eff}$ depends on the values of $\alpha$ and $\beta$. The specific form of the equation allows for a smooth transition between the two phases, and its parameters can be constrained by fitting to cosmological observations.

From Eqs. (\ref{F1}) and (\ref{F2}), one can derive
\begin{equation}
    \frac{\overset{..}{a}}{a}=-\frac{1}{6}\left( \rho _{m}+\rho
_{DE}+3p_{DE}\right).
\end{equation}

The present $\omega_{eff}$ model will result in acceleration ($\overset{..}{a}>0$) only if the following condition is satisfied,
\begin{equation}
    \omega_{eff}=\frac{p_{DE}}{\rho_{m}+\rho_{DE}}<-\frac{1}{3}.
\end{equation}

Using the ansatz for $\omega_{eff}(z)$ as given in Eq. (\ref{eff}), we can obtain an additional constraint on the values of the parameters $\alpha$ and $\beta$ as $2\alpha>\beta$. 

Again, by introducing this ansatz in Eqs. (\ref{omegaH}) and (\ref{t_z}), we obtain the differential equation for the Hubble parameter $H$. This equation can be expressed as follows:
\begin{equation}
    -1+\frac{2}{3}(1+z)\frac{1}{H(z)}\frac{dH(z)}{dz}=-\frac{\alpha }{\alpha +(z+1) (\beta +z)}.
\end{equation}

The cosmological solution for the Hubble parameter $H(z)$ as a function of redshift is obtained as,
\begin{widetext}
\begin{equation}
    H(z)=H_{0}\frac{(\alpha +\beta +z (\beta +z+1))^{3/4} \exp \left(-\frac{3 (\beta -1) \left(\tan ^{-1}\left(\frac{\beta +1}{\sqrt{4 \alpha -(\beta -1)^2}}\right)-\tan ^{-1}\left(\frac{\beta +2 z+1}{\sqrt{4 \alpha -(\beta -1)^2}}\right)\right)}{2 \sqrt{4 \alpha -(\beta -1)^2}}\right)}{(\alpha +\beta )^{3/4}}.
    \label{Hz}
\end{equation}
\end{widetext}

The expression for the Hubble parameter in terms of redshift, given by Eq. (\ref{Hz}), includes a constant term $H_{0}$, which represents the present value of the Hubble parameter. This constant term reflects the current rate of expansion of the Universe and is an important quantity in cosmology. The value of $H_{0}$ has been a topic of much research and debate, with different methods yielding slightly different values. However, it is currently accepted that the most precise determination of $H_{0}$ comes from the Planck measurements, which give a value of $H_{0} = 67.4 \pm 0.5 km/s/Mpc$ \cite{Planck2020}. The $\Lambda$CDM model has a constant EoS parameter for DE, given by $\omega_{\Lambda}=-1$. This corresponds to a constant Hubble parameter as a function of redshift, given by:
\begin{equation}
H(z)=H_0 \sqrt{\Omega_{m0}(1+z)^3 + \Omega_{\Lambda0}},
\end{equation}
where $\Omega_{m0}=\frac{\rho_{m0}}{3H_{0}^2}$ and $\Omega_{\Lambda0}=\frac{\rho_{\Lambda0}}{3H_{0}^2}$ are the present-day density parameters for matter and DE, respectively.

Since the equations that describe the evolution of the Universe are highly complex and lengthy. Here, we will write only the general definition of some important cosmological parameters in this context. 

The density parameters describe the fractional contribution of each component to the total energy density of the Universe at a given time. They are defined as the ratio of the energy density of a particular component to the critical energy density of the Universe required to achieve a flat geometry. The critical energy density is given by the expression $3H^2$. The values of the density parameters for DE and matter can be calculated using the relations
\begin{equation}
    \Omega_{m}\left( z\right) =\frac{\rho_{m}}{3H^{2}}, 
\end{equation}
\begin{equation}
    \Omega_{DE}\left( z\right) =\frac{\rho_{DE}}{3H^{2}}.
\end{equation}

Finally, the DE EoS parameter describes the relationship between the pressure and energy density of the DE component. It is defined as the ratio of the pressure of DE to its energy density,
\begin{equation}
    \omega _{DE}(z)=\frac{p_{DE}}{\rho_{DE}}=-\frac{\left( 2\overset{.}{H}+3H^{2}\right) }{3H^{2}-\rho_{m}}.
\end{equation}

The expanded expressions of the density parameters and DE EoS parameter for the present $\omega_{eff}$ model are provided in Appendix \ref{app}.

\section{Statistical Analysis Methods}
\label{sec3}

To assess the validity and accuracy of the proposed model, we employ observational data obtained from three distinct datasets: the Cosmic Chronometer (CC) dataset, the Baryon Acoustic Oscillations (BAO) dataset, and the Type Ia Supernova (SN) dataset. This section outlines the methodology we used to incorporate these datasets and extract relevant information. This topic has also been presented in a similar manner by M. Koussour et al. \cite{fQ6,fQ7}, who have also discussed the observational data in detail.

In this study, we first utilized a Bayesian statistical analysis technique, along with the \textit{emcee} python library \cite{Mackey/2013}, to carry out a Markov chain Monte Carlo (MCMC) simulation. This simulation allowed us to derive the posterior probability distribution of the model parameters. The posterior probability distribution provides a probability distribution that represents our level of confidence in the values of the model parameters after taking into account the available observational data. Once the MCMC simulation was completed, we analyzed the resulting chain of parameter values to determine the best-fit values and the uncertainties associated with the model parameters. The parameter space of our model can be defined as $\theta_{s}=(H_{0}, \alpha, \beta)$. The best-fit values for the parameters can be determined using the probability function $\mathcal{L} \propto exp(-\chi^2/2)$, where $\chi ^{2}$ represents the chi-squared function. This statistical tool is commonly used in cosmology to obtain the parameters of a specific cosmological model that best fit observed data. The aim is to find the values of $\theta_{s}$ that result in the minimum value of $\chi^{2}$. Typically, an MCMC approach is employed to explore the parameter space and identify the regions that have the highest likelihood given the observational data.

\subsection{Cosmic Chronometer (CC) dataset}

In the present work, we use data obtained from the CC dataset, which consists of Hubble parameter measurements obtained using the differential age (DA) method. In particular, we utilize 31 data points obtained from Refs. \cite{Yu/2018, Moresco/2015, Sharov/2018} that provide measurements of the Hubble parameter at different redshifts. These measurements are crucial for constraining the Universe's expansion history and testing different cosmological models. To quantify how well our model fits the data, we use the $\chi ^{2}$ function, which is defined as follows:
\begin{equation}
\chi^{2}_{CC} = \sum_{i=1}^{31} \frac{\left[H(\theta_{s}, z_{i})-
H_{obs}(z_{i})\right]^2}{\sigma(z_{i})^2}.
\end{equation}

In this equation, $H(\theta_s, z_i)$ is the theoretical value of the Hubble parameter at redshift $z_i$ for a particular set of cosmological parameters $\theta_s$, $H_{obs}(z_i)$ is the measured value of the Hubble parameter at redshift $z_i$, and $\sigma (z_{i})$ is the corresponding uncertainty of $H_{i}$.

\subsection{Baryon Acoustic Oscillations (BAO) dataset}

Next, we consider the BAO dataset obtained from multiple surveys, such as the 6dFGS (Six Degree Field Galaxy Survey), the SDSS (Sloan Digital Sky Survey), and the LOWZ samples of the BOSS (Baryon Oscillation Spectroscopic Survey) \cite{BAO1,BAO2,BAO3,BAO4,BAO5,BAO6}. These surveys have provided highly precise measurements of the positions of the BAO peaks in galaxy clustering at different redshifts. The characteristic scale of BAO can be determined by the sound horizon $r_s$ at the epoch of photon decoupling with redshift $z_{dec}$, which is related by the following equation:
\begin{equation}\label{4b}
r_{s}(z_{\ast })=\frac{c}{\sqrt{3}}\int_{0}^{\frac{1}{1+z_{\ast }}}\frac{da}{
a^{2}H(a)\sqrt{1+(3\Omega _{b,0}/4\Omega _{\gamma,0})a}},
\end{equation}
where $\Omega _{b,0}$ and $\Omega _{\gamma,0}$ denote the present densities values of Baryon and photon respectively. The BAO dataset utilized in this study consists of six data points for $d_{A}(z_{\ast })/D_{V}(z_{BAO})$, which were obtained from the sources listed in Refs. \cite{BAO1, BAO2, BAO3, BAO4, BAO5, BAO6}. Here, $z_{\ast }\approx 1091$ denotes the redshift value for photon decoupling, and $d_{A}(z_{\ast })=c\int_{0}^{z}\frac{dz'}{H(z')}$ denotes the comoving angular diameter distance at decoupling, along with the dilation scale $D_{V}(z)=\left[czd_{A}^{2}(z)/H(z)\right] ^{1/3}$. 

The chi-square function presented in \cite{BAO6} is utilized to evaluate the BAO dataset, and it is expressed as,
\begin{equation}\label{4e}
\chi _{BAO}^{2}=X^{T}C_{BAO}^{-1}X,
\end{equation}
where 
\begin{equation}
X=\left( 
\begin{array}{c}
\frac{d_{A}(z_{\star })}{D_{V}(0.106)}-30.95 \\ 
\frac{d_{A}(z_{\star })}{D_{V}(0.2)}-17.55 \\ 
\frac{d_{A}(z_{\star })}{D_{V}(0.35)}-10.11 \\ 
\frac{d_{A}(z_{\star })}{D_{V}(0.44)}-8.44 \\ 
\frac{d_{A}(z_{\star })}{D_{V}(0.6)}-6.69 \\ 
\frac{d_{A}(z_{\star })}{D_{V}(0.73)}-5.45%
\end{array}%
\right) \,,
\end{equation}%
and $C_{BAO}^{-1}$ is the inverse of the covariance matrix \cite{BAO6}.

\subsection{Type Ia Supernovae (SN) dataset}

The data from SN observations is a common data sample used to analyze the late-time behavior of the Universe. The distance modulus of SN is defined as the difference between the observed apparent magnitude $m_{B}$ and the absolute magnitude $M_{B}$ of the $B$ band spectrum,
\begin{equation}
\mu(z)=5log_{10}\frac{d_{L}(z)}{1Mpc}+25,
\end{equation}%
where the $d_{L}(z)$ is the luminosity distance and in a spatially flat FLRW Universe is defined as,
\begin{equation}
d_{L}(z)=c(1+z)\int_{0}^{z}\frac{dz'}{H(z',\theta_{s} )},
\end{equation}%

In the present work, we consider the Pantheon sample, which is constructed from the PanSTARRS1 (Panoramic Survey Telescope and Rapid Response System) Medium Deep Survey, SDSS (Sloan Digital Sky Survey), SNLS (Supernova Legacy Survey), and numerous low-z, and HST (Hubble Space Telescope) samples, comprises 1048 data points covering a broad range of redshifts $0.01\leq z \leq2.26$ and is a valuable dataset of SN \cite{Scolnic/2018,Chang/2019}. 

The $\chi ^{2}$ function for SNe dataset is expressed as,
\begin{equation}
\chi _{SN}^{2}=\sum_{i,j=1}^{1048}\Delta \mu _{i}\left( C_{SN}^{-1}\right)
_{ij}\Delta \mu _{j},
\end{equation}%
where $\Delta \mu_{i}=\mu_{\rm th}-\mu_{\rm obs}$ is the difference between the theoretical and observed distance modulus, and $C_{SN}^{-1}$ is the inverse of the covariance matrix of the Pantheon sample.

\subsection{CC+BAO+SN dataset}

Now, the maximum likelihood method can be employed by taking the total likelihood function as the product of the likelihoods of individual datasets,
\begin{equation} 
\mathcal{L}_{Joint}=\mathcal{L}_{CC}\times \mathcal{L}_{BAO}\times \mathcal{L}_{SN},
\end{equation}
where
\begin{equation}
\chi _{Joint}^{2} =\chi _{CC}^{2}+\chi _{BAO}^{2}+\chi _{SN}^{2}.
\end{equation}

\begin{widetext}

\begin{figure}[h]
\centerline{\includegraphics[scale=0.8]{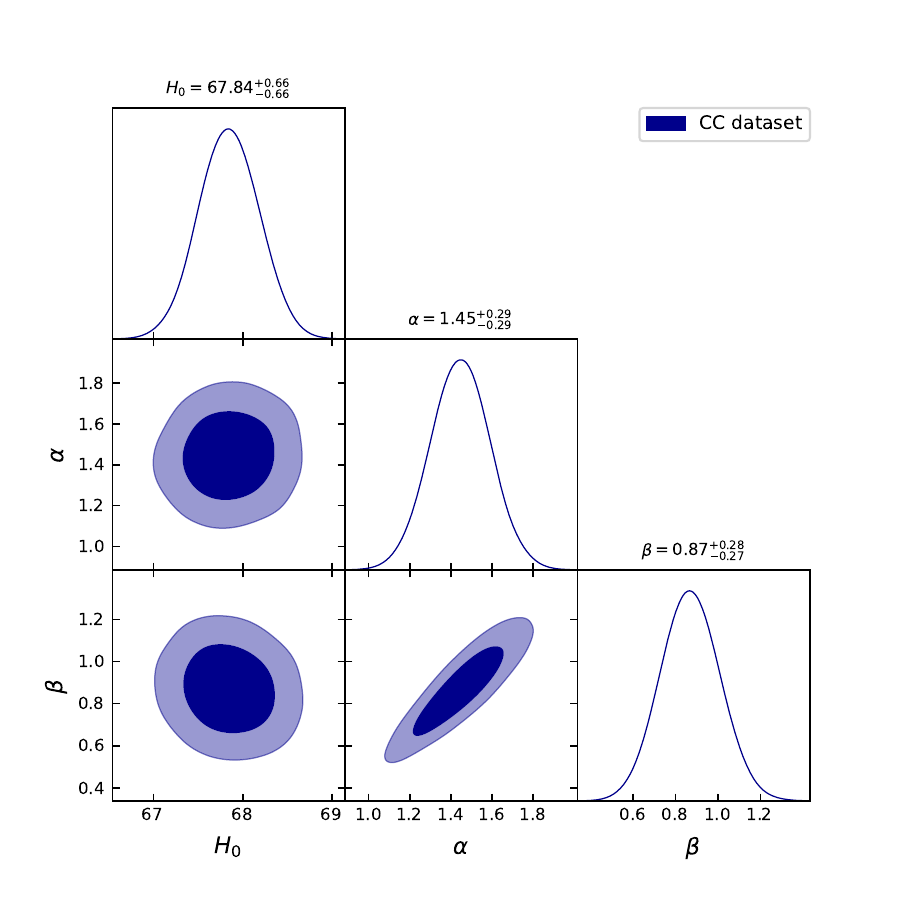}}
\caption{The plot displays the likelihood contours at $1-\sigma$ and $2-\sigma$ CL on the parameters $H_{0}$, $\alpha$, and $\beta$ for the present $\omega_{eff}$ model using CC dataset. The dark blue and light blue shaded regions represent $1-\sigma$ and $2-\sigma$ CL, respectively. }
\label{CC}
\end{figure}

\begin{figure}[h]
\centerline{\includegraphics[scale=0.8]{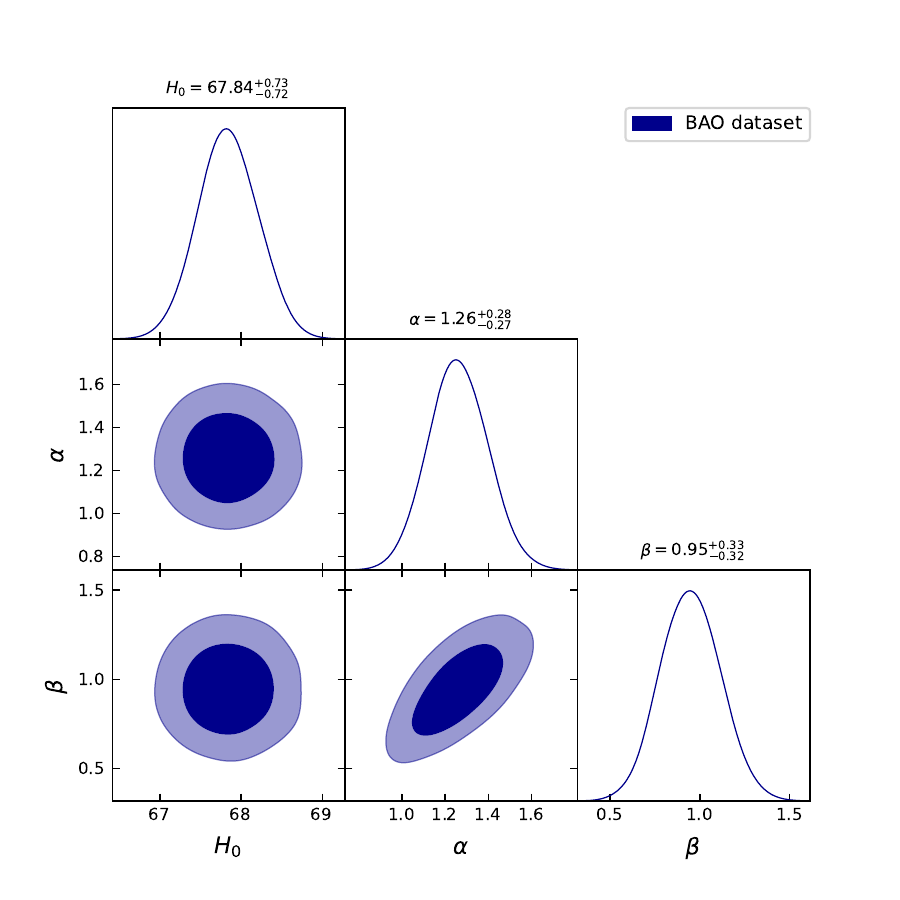}}
\caption{The plot displays the likelihood contours at $1-\sigma$ and $2-\sigma$ CL on the parameters $H_{0}$, $\alpha$, and $\beta$ for the present $\omega_{eff}$ model using BAO dataset. The dark blue and light blue shaded regions represent $1-\sigma$ and $2-\sigma$ CL, respectively.}
\label{CC+BAO}
\end{figure}

\begin{figure}[h]
\centerline{\includegraphics[scale=0.8]{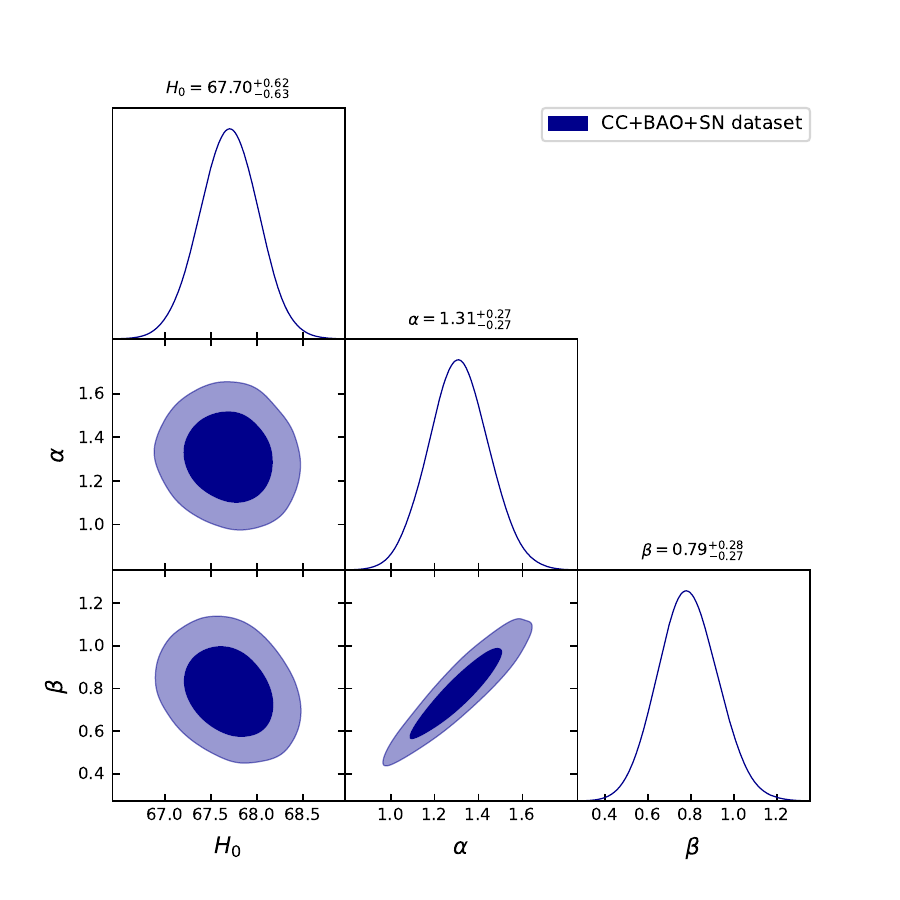}}
\caption{The plot displays the likelihood contours at $1-\sigma$ and $2-\sigma$ CL on the parameters $H_{0}$, $\alpha$, and $\beta$ for the present $\omega_{eff}$ model using CC+BAO+SN dataset. The dark blue and light blue shaded regions represent $1-\sigma$ and $2-\sigma$ CL, respectively.}
\label{CC+BAO+SN}
\end{figure}

\begin{figure}[h]
\centerline{\includegraphics[scale=0.60]{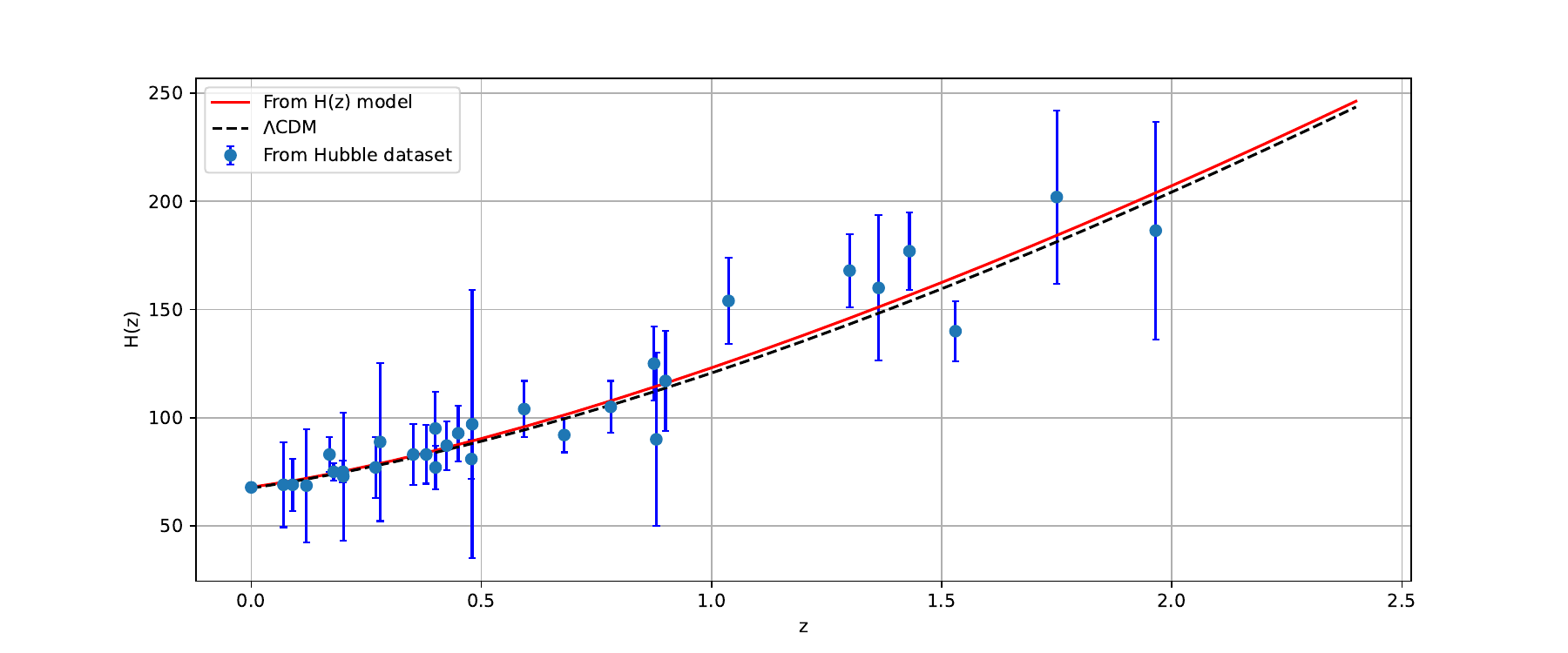}}
\caption{The plot displays a comparison between the present $\omega_{eff}$ model and the $\Lambda$CDM model in terms of the Hubble parameter $H(z)$ against redshift $z$. The present $\omega_{eff}$ model is represented by the red line while the black dotted line represents the $\Lambda$CDM model. The 31 data points from the CC dataset are displayed on the plot along with their error bars, which our model fits well.}
\label{ErrorHubble}
\end{figure}

\begin{figure}[h]
\centerline{\includegraphics[scale=0.60]{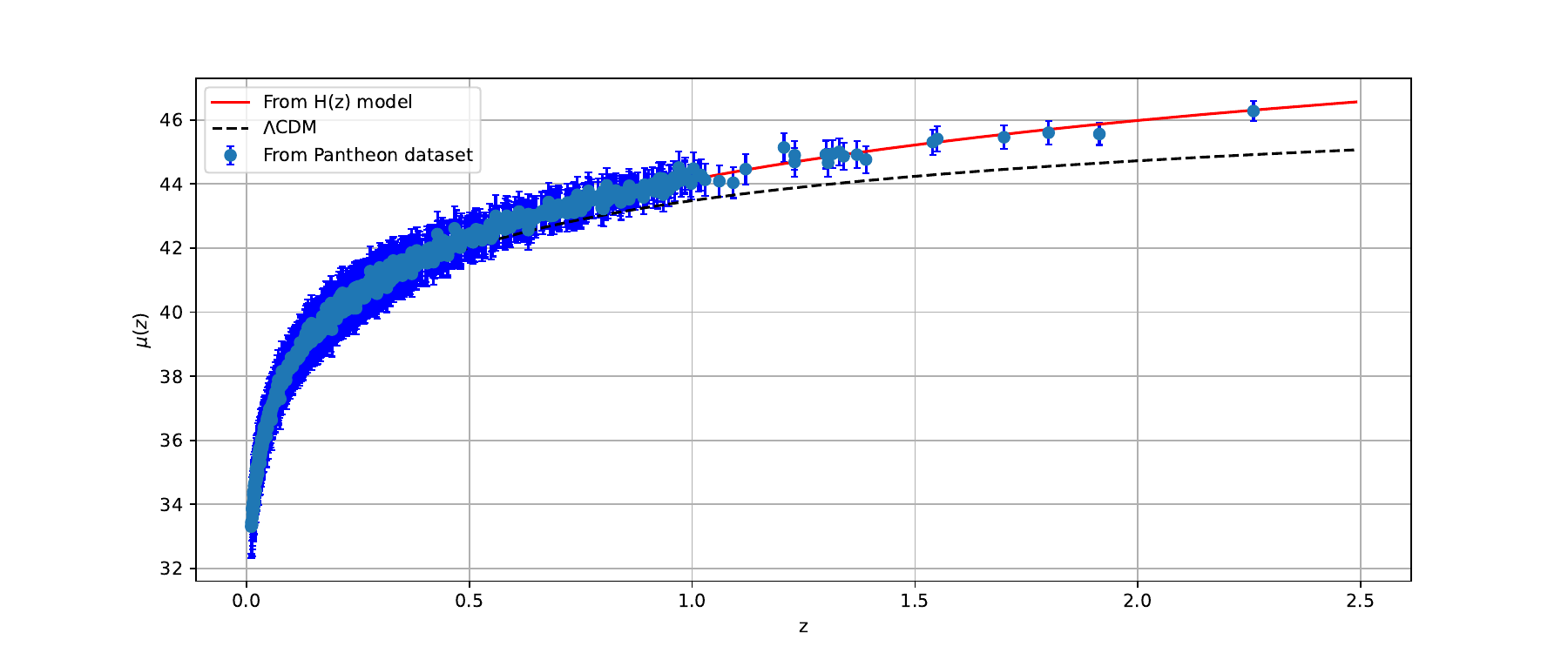}}
\caption{The plot displays a comparison between the present $\omega_{eff}$ model and the $\Lambda$CDM model in terms of the distance modulus $\mu(z)$ against redshift $z$. The present $\omega_{eff}$ model is represented by the red line while the black dotted line represents the $\Lambda$CDM model. The 1048 data points of the SNe dataset are displayed on the plot along with their error bars, which our model fits well.}
\label{ErrorSNe}
\end{figure}

\begin{table*}[h]
\begin{center}
\renewcommand{\arraystretch}{1.5}
\begin{tabular}{l c c c c c c c c c}
\hline 
Dataset  & $H_{0}(km \, s^{-1}\, Mpc^{-1})$ & $\alpha$ & $\beta$ & $q_{0}$ & $z_{tr}$ & $\omega _{0}$\\
\hline
$Priors$ & $(60,80)$ & $(-10,10)$ & $(-10,10)$ & $-$ & $-$ & $-$\\

$CC$   & $67.84_{-0.66}^{+0.66}$  & $1.45_{-0.29}^{+0.29}$ & $0.87_{-0.27}^{+0.28}$ & $-0.44_{-0.03}^{+0.04}$ & $0.77_{-0.03}^{+0.02}$ & $-0.91_{-0.02}^{+0.01}$\\

$BAO$   & $67.84_{-0.72}^{+0.73}$  & $1.26_{-0.27}^{+0.28}$ & $0.95_{-0.27}^{+0.28}$ & $-0.36_{-0.04}^{+0.04}$ & $0.61_{-0.01}^{+0.01}$ & $-0.83_{-0.02}^{+0.02}$\\

$CC+BAO+SN$   & $67.70_{-0.63}^{+0.62}$  & $1.31_{-0.27}^{+0.27}$ & $0.79_{-0.27}^{+0.28}$ & $-0.44_{-0.04}^{+0.05}$ & $0.73_{-0.02}^{+0.01}$ & $-0.91_{-0.03}^{+0.03}$\\
\hline
\end{tabular}
\caption{The table displays the marginalized constraints on the parameters $H_{0}$, $\alpha$, and $\beta$ for the CC, BAO, and CC+BAO+SN datasets, with a 68\% CL. }
\label{tab1}
\end{center}
\end{table*}
\end{widetext}

\section{Statistical Analysis Results}
\label{sec5}

In the present work, the values of the model parameters $\theta_{s}=(H_{0}, \alpha, \beta)$ have been estimated using the $\chi^{2}$ minimization technique, which is equivalent to the maximum likelihood analysis. Moreover, we have generated two-dimensional likelihood contours with $1-\sigma$ and $2-\sigma$ errors, corresponding to the 68\% and 95\% confidence levels (CL), respectively, for three different data samples: CC, BAO, and CC+BAO+SN.  These likelihood contours are shown in Figs. \ref{CC}, \ref{CC+BAO}, and \ref{CC+BAO+SN}. The likelihood plots show that the Gaussian distribution fits the likelihood functions well for three datasets. For the CC dataset consisting of 31 data points, we obtained the value of $\alpha$ to be $1.45_{-0.29}^{+0.29}$, and the constrained value for $\beta$ was $0.87_{-0.27}^{+0.28}$. For the BAO dataset with six sample points, we obtained the values of $\alpha$ and $\beta$ to be $1.26_{-0.27}^{+0.28}$ and $0.95_{-0.27}^{+0.28}$, respectively. Finally, for the combined dataset in the last section, consisting of CC, BAO, and SN datasets, we obtained the values of $\alpha$ and $\beta$ to be $1.31_{-0.27}^{+0.27}$ and $0.79_{-0.27}^{+0.28}$, respectively. Now, for the present value of the Hubble parameter, we found $H_{0}=67.84_{-0.66}^{+0.66}$, $H_{0}=67.84_{-0.72}^{+0.73}$, $H_{0}=67.70_{-0.63}^{+0.62}$ for CC,
BAO, and CC+BAO+SN datasets, respectively \cite{Chen1,Chen2,Aubourg}. In addition to the estimation of the parameters, we compared the performance of our model with the standard $\Lambda$CDM model using the evolution of the Hubble parameter $H(z)$ and distance modulus $\mu(z)$. The constraint values of the model parameters $\theta_{s}=(H_{0}, \alpha, \beta)$ for the CC and SN samples were used for this comparison, and the results are shown in Figs. \ref{ErrorHubble} and \ref{ErrorSNe}. The plots indicate that our model is in good agreement with the observational data for both CC and SN samples. Furthermore, the comparison shows that our model is very close to the evolution of the $\Lambda$CDM model. The statistical analysis results for the present $\omega_{eff}$ model are summarized in Tab. \ref{tab1}.

\begin{figure}[h]
\centerline{\includegraphics[scale=0.68]{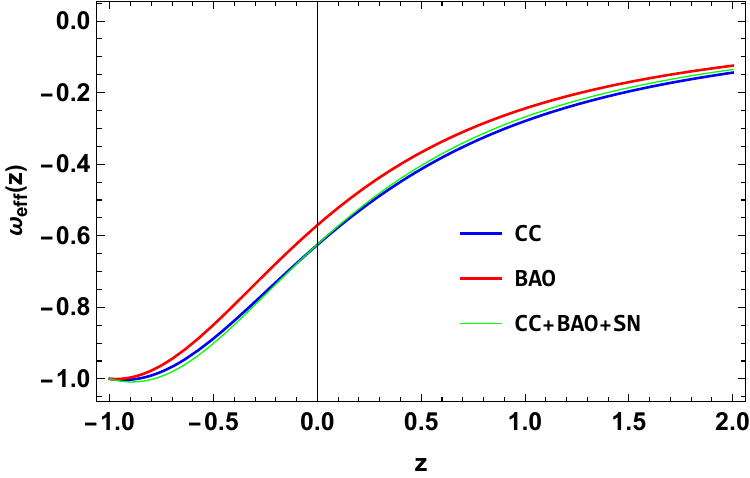}}
\caption{The plot displays the evolution of the effective EoS parameter $\omega_{eff}$ with respect to the redshift $z$ using the values constrained from the CC, BAO, and CC+BAO+SN datasets.}
\label{F_EoS_eff}
\end{figure}

\begin{figure}[h]
\centerline{\includegraphics[scale=0.68]{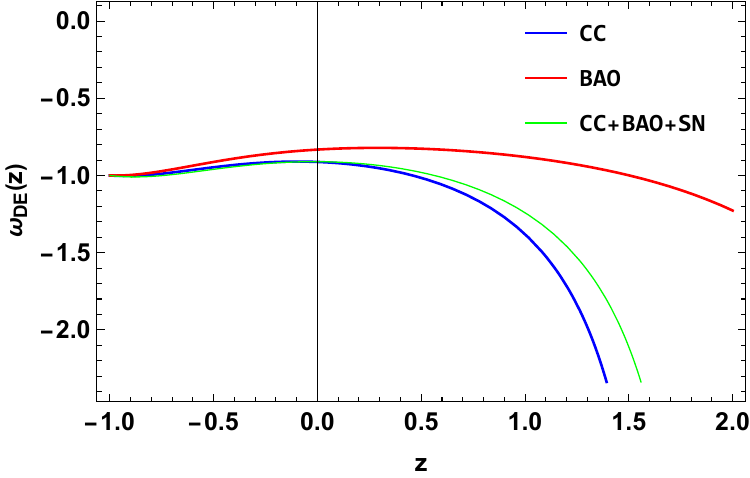}}
\caption{The plot displays the evolution of the DE EoS parameter $\omega_{DE}$ with respect to the redshift $z$ using the values constrained from the CC, BAO, and CC+BAO+SN datasets.}
\label{F_EoS_DE}
\end{figure}

Furthermore, it is generally known that the EoS parameter $\omega$ is a key factor in describing the energy-dominated evolution process of the Universe, and it is known that it can predict the present scenario of the Universe in either the quintessence phase ($-1<\omega<-\frac{1}{3}$) or the phantom phase ($\omega<-1$). The acceleration of both the effective EoS parameter and the EoS parameter for the DE is shown in Figs. \ref{F_EoS_eff} and \ref{F_EoS_DE} for the CC, BAO, and CC+BAO+SN dataset. The effective EoS parameter begins in the matter-dominated phase and then transitions through the quintessence phase before finally reaching a constant value in the cosmological constant-dominated region. Similarly, the EoS parameter for DE initially displays phantom-like behavior in the early Universe, then moves to the quintessence phase in the present, and eventually transitions to the cosmological constant phase. This behavior of the EoS parameter for DE is consistent with the behavior of the effective EoS parameter in the future. Also, it is observable that in the future, both EoS behavior has a minor shift toward the phantom region before it asymptotically approaches the cosmological constant. In this study, we have obtained the values of $\omega_{0}$ for the CC, BAO, and CC+BAO+SN dataset as $\omega_{0}=-0.91_{-0.02}^{+0.01}$, $\omega_{0}=-0.83_{-0.02}^{+0.02}$, and $\omega_{0}=-0.91_{-0.03}^{+0.03}$, respectively \cite{Hernandez,Zhang}. These results highlight the importance of the EoS parameter in understanding the evolution of the Universe and suggest that the present model is consistent with the quintessence phase.

\begin{figure}[h]
\centerline{\includegraphics[scale=0.65]{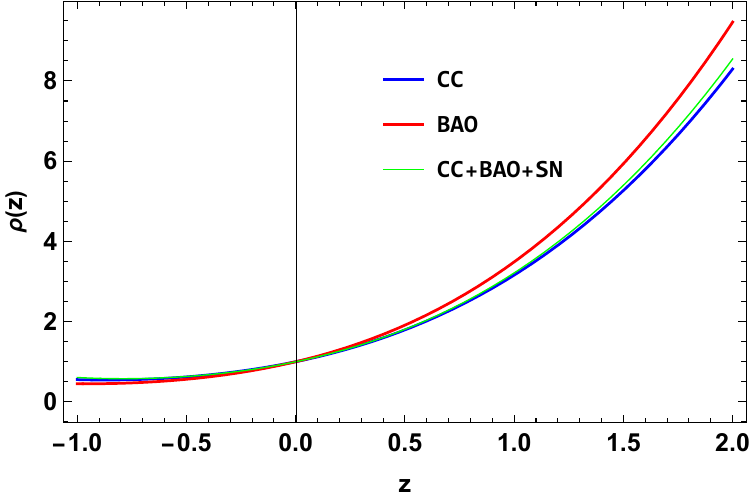}}
\caption{The plot displays the evolution of the energy density $\rho$ with respect to the redshift $z$ using the values constrained from the CC, BAO, and CC+BAO+SN datasets.}
\label{F_rho}
\end{figure}

In Fig. \ref{F_rho}, we can observe that the energy density behaves as expected, showing a positive value and decreasing with the expansion of the Universe in both the present and far future. The plot also indicates that the energy density of DE dominates over the matter density at late times. Moreover, the behavior of the energy density plot is in agreement with the predictions of several theoretical models of DE.

\begin{figure}[h]
\centerline{\includegraphics[scale=0.65]{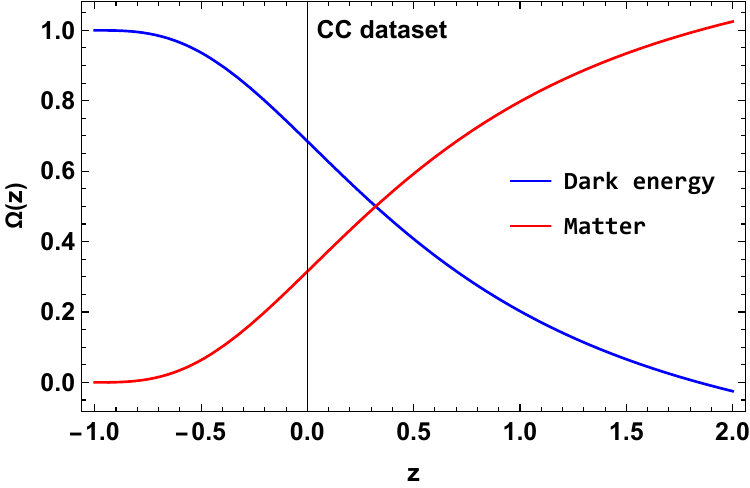}}
\caption{The plot displays the evolution of the density parameters $\Omega_{DE}$ and $\Omega_{m}$ with respect to the redshift $z$ using the values constrained from the CC dataset.}
\label{F_Omega1}
\end{figure}

\begin{figure}[h]
\centerline{\includegraphics[scale=0.65]{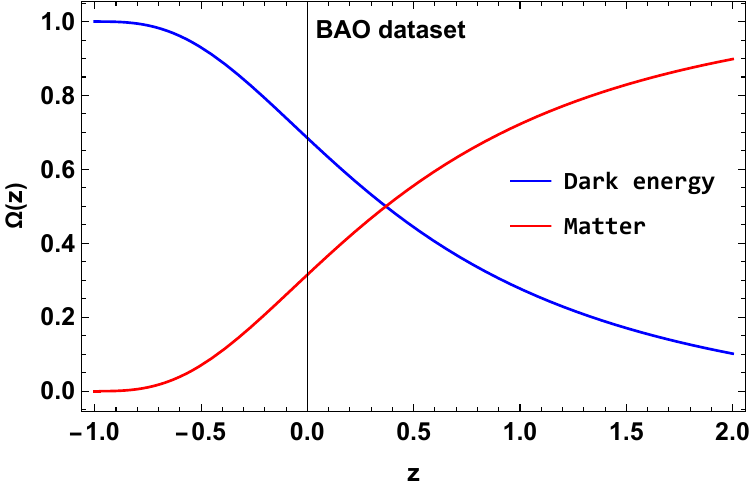}}
\caption{The plot displays the evolution of the density parameters $\Omega_{DE}$ and $\Omega_{m}$ with respect to the redshift $z$ using the values constrained from the BAO dataset.}
\label{F_Omega2}
\end{figure}

\begin{figure}[h]
\centerline{\includegraphics[scale=0.65]{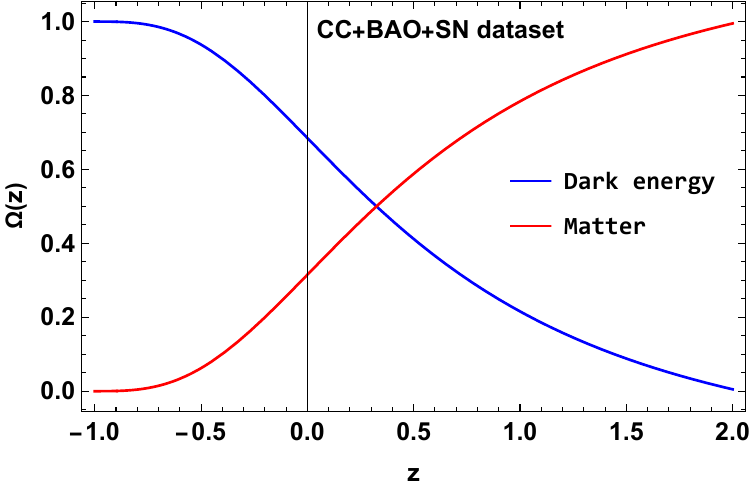}}
\caption{The plot displays the evolution of the density parameters $\Omega_{DE}$ and $\Omega_{m}$ with respect to the redshift $z$ using the values constrained from the CC+BAO+SN dataset.}
\label{F_Omega3}
\end{figure}

In Figs. \ref{F_Omega1}, \ref{F_Omega2}, and \ref{F_Omega3}, we can observe the evolution of the density parameter for matter and DE for the model parameters constrained by the CC, BAO, and CC+BAO+SN datasets, respectively. In the early period, the Universe is mostly dominated by matter, while the DE density parameter remains negligible. However, as the Universe expands, the matter density parameter gradually decreases due to the increase in volume, while the DE density parameter becomes dominant. This eventually leads to the acceleration of the Universe's expansion, which is consistent with the late-time cosmic acceleration of the Universe observed through various cosmological surveys. The consistency of the observed behavior is particularly significant given that $\Omega_{m0}=0.315 \pm0.007$, as determined by the latest Planck data \cite{Planck2020}, was used in all three of the aforementioned figures. The behavior of the density parameter for DE indicates the presence of an unknown energy component that drives the current phase of accelerated expansion. 

\section{Investigation of Geometric Parameters of the Model}
\label{sec4}

\subsection{Deceleration parameter}
The deceleration parameter is another key cosmological parameter that describes the expansion rate of the Universe. It is defined as
\begin{equation}
    q = -\frac{1}{H^2}\frac{\ddot{a}}{a}=-1- \frac{\dot{H}}{H^2}.
\end{equation}

 In terms of the Hubble parameter $H$ and its derivative with respect to redshift $z$, the deceleration parameter can be expressed as,
 \begin{equation}
     q(z) =-1+ \frac{1+z}{H(z)} \frac{d}{dz} \left[ H(z) \right].
     \label{q}
 \end{equation}

For the present $\omega_{eff}$ model, using Eqs. (\ref{Hz}) and (\ref{q}), $q(z)$ takes the following form
\begin{equation}
    q(z)=\frac{1}{2}-\frac{3 \alpha }{2 (\alpha +\beta +z (\beta+1+z))}.
    \label{qz}
\end{equation}

Understanding the value of the deceleration parameter is crucial in determining the fate of the Universe. If $q > 0$, the Universe is decelerating, meaning the expansion is slowing down over time. On the other hand, if $q < 0$, the Universe is accelerating, and the expansion is speeding up over time.  If $q$ is equal to $-1$, it corresponds to an exponential expansion, also known as the de Sitter (dS) expansion. If $q$ is less than $-1$, it indicates a superexponential expansion, whereas if $-1 < q < 0$, it represents a power-law accelerating rate. These scenarios are of great interest in cosmology as they help us to understand the behavior of DE and its effect on the expansion of the Universe.

\begin{figure}[h]
\centerline{\includegraphics[scale=0.68]{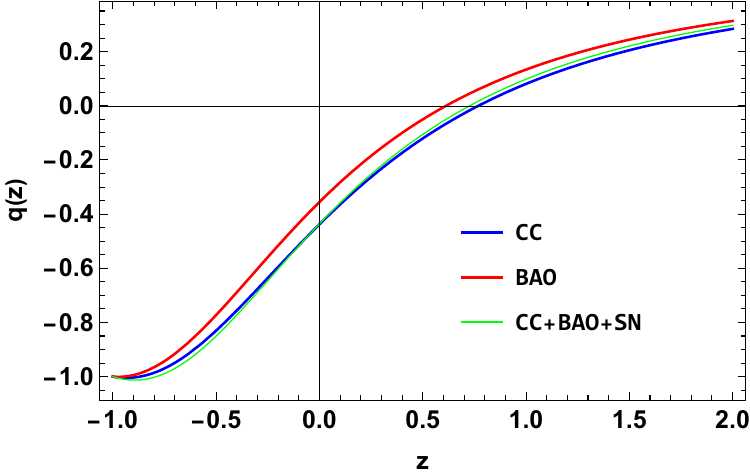}}
\caption{The plot displays the evolution of the deceleration parameter $q$ with respect to the redshift $z$ using the values constrained from the CC, BAO, and CC+BAO+SN datasets.}
\label{F_q}
\end{figure}

The results depicted in Fig. \ref{F_q} demonstrate a remarkable similarity between the evolution of the effective EoS $\omega_{eff}$ and the deceleration parameter $q(z)$ for the proposed model. Moreover, the deceleration parameter exhibits a significant change in its behavior around the redshift values of 0.5 and 0.8, known as a signature flip. This signature flip is in excellent agreement with the analysis of observational data conducted by \cite{Farooq0}. Also, This plot presents a clear depiction of the past decelerating expansion of the Universe and the current acceleration in its late-time evolution. The present values of the deceleration parameter are $q_{0}=-0.44_{-0.03}^{+0.04}$, $q_{0}=-0.36_{-0.04}^{+0.04}$, $q_{0}=-0.44_{-0.04}^{+0.05}$ corresponding to the model parameters constrained by the CC,
BAO, and CC+BAO+SN datasets, respectively. 

Further, the transition redshift, denoted as $z_{tr}$, can be determined by finding the redshift at which the deceleration parameter $q(z_{tr})=0$, 
\begin{equation}
z_{tr} =\frac{1}{2} \left(\sqrt{8 \alpha +\beta ^2-2 \beta +1}-\beta -1\right).
\end{equation}

By fitting the model parameters to match the observational data, we determine the following values of the transition redshift $z_{tr}$ as $z_{tr}=0.77_{-0.03}^{+0.02}$, $z_{tr}=0.61_{-0.01}^{+0.01}$, $z_{tr}=0.73_{-0.02}^{+0.01}$ for the CC, BAO, and CC+BAO+SN datasets, respectively \cite{Farooq,Jesus,Garza}. Therefore, these findings provide strong support for the proposed model's validity in describing the dynamics of the Universe's expansion.

\subsection{State finder diagnostics}
Several dynamical models of DE have been introduced to overcome the issues faced by the concept of a cosmological constant $\Lambda$, such as the fine-tuning problem and the cosmological coincidence problem, as previously discussed in the introduction. It is important to distinguish between these time-varying DE models and determine which one best fits observational data. For this reason, V. Sahni et al. \cite{Sahni, Alam} introduced state finder parameters ($r$, $s$) as a new pair of geometrical parameters. These parameters have become a popular tool in modern cosmology and provide a powerful means of discriminating between different models of DE. The state finder parameters are defined as,
\begin{equation}
r=\frac{\overset{...}{a}}{aH^{3}}=2q^{2}+q-\frac{\overset{.}{q}}{H},  
\end{equation}%
\begin{equation}
s=\frac{\left( r-1\right) }{3\left( q-\frac{1}{2}\right) }.
\end{equation}

Using the expressions for the Hubble parameter (Eq. (\ref{Hz})) and the deceleration parameter (Eq. (\ref{qz})), we can calculate the state finder parameters for the present $\omega_{eff}$ model as,
\begin{equation}
r(z)=1-\frac{3 \alpha  (1+z) (2 \beta +z-1)}{2 (\alpha +\beta +z (\beta +1+z))^2},  
\end{equation}%
\begin{equation}
s(z)=\frac{(1+z) (2 \beta +z-1)}{3 (\alpha +\beta +z (\beta +1+z))}.
\end{equation}

\begin{figure}[h]
\centerline{\includegraphics[scale=0.73]{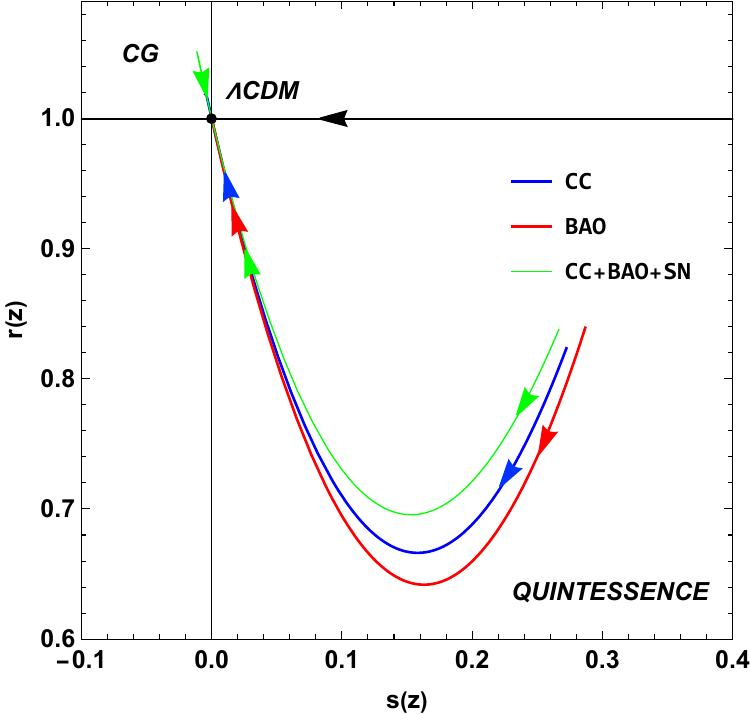}}
\caption{The plot displays the evolution of the $r-s$ plane using the values constrained from the CC, BAO, and CC+BAO+SN datasets with $-1\leq z\leq2$.}
\label{F_rs}
\end{figure}

\begin{figure}[h]
\centerline{\includegraphics[scale=0.73]{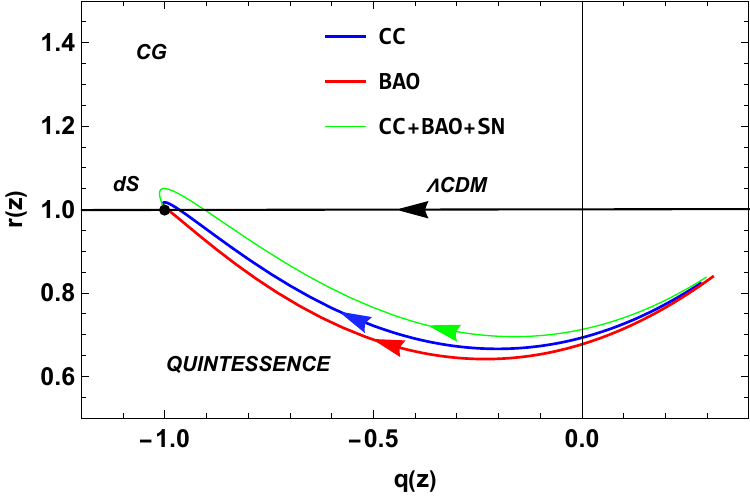}}
\caption{The plot displays the evolution of the $r-q$ plane using the values constrained from the CC, BAO, and CC+BAO+SN datasets with $-1\leq z\leq2$.}
\label{F_rq}
\end{figure}

Different DE models have distinct trajectories in the $r-s$ plane, which can be used to differentiate between them. The state finder parameters, denoted as ($r$, $s$), can take different values depending on the DE model. The values corresponding to different DE models are as follows: the $\Lambda$CDM model is analogous to $(r=1,s=0)$, the SCDM (Standard Cold Dark Matter) model is analogous to $(r=1,s=1)$, the HDE (Holographic DE) model is analogous to $(r=1,s=\frac{2}{3})$, the Chaplygin Gas (CG) model is analogous to $(r>1,s<0)$, and the Quintessence model is analogous to $(r<1,s>0)$. Therefore, by observing the values of $r$ and $s$, one can determine which DE model is consistent with the observational data. In the $r-s$ plane depicted in Fig. \ref{F_rs}, it is evident that the present $\omega_{eff}$ model initially has values of $r<1$ and $s>0$, which indicates that the DE behaves similarly to quintessence, with its energy density decreasing as the Universe expands. However, with time, the model gradually converges to the $\Lambda$CDM model, represented by $r=1$ and $s=0$. Similarly, the $r-q$ plane illustrated in Fig. \ref{F_rq} suggests that the current state of the Universe in the model is dominated by a quintessential fluid. Still, it is expected to move towards a dS phase ($r=1$ and $q=-1$), where the Universe is dominated by a cosmological constant, leading to a constant rate of expansion. The observed behavior of the state finder parameters for the constrained values of the the model parameters from the CC, BAO, and CC+BAO+SN datasets is consistent with the behavior of the cosmological parameters explained in the preceding section. The model exhibits quintessence-like behavior initially but eventually transitions to a $\Lambda$CDM-like behavior, implying that the DE in the Universe varies over time. This is an important finding as it highlights that the state finder parameters serve as useful tools for distinguishing between different DE models and understanding the Universe's evolution.

\subsection{$Om(z)$ diagnostic}
The $Om(z)$ diagnostic, which is another useful method for distinguishing between different DE models in cosmology \cite{Sahni1}, involves only the Hubble parameter and is simpler than the state finder parameters. The Hubble parameter is obtained by taking the first derivative of the scale factor of the Universe. In the case of a spatially flat Universe, the $Om(z)$ diagnostic is defined as,
\begin{equation}
Om(z) = \frac{\left(\frac{H}{H_{0}}\right)^{2}-1}{(1+z)^{3}-1}.
\label{Omz}
\end{equation}

For the present $\omega_{eff}$ model, by using Eqs. (\ref{Hz}) and (\ref{Omz}), the expression for the $Om(z)$ diagnostic is derived and presented in the appendix \ref{app}.

The slope of $Om(z)$ can provide insights into the behavior of DE, where a negative slope corresponds to quintessence behavior ($\omega>-1$), and a positive slope corresponds to phantom-type behavior ($\omega<-1$). On the other hand, a constant $Om(z)$ indicates the $\Lambda$CDM model, where DE is represented by a cosmological constant. The $Om(z)$ diagnostic can be employed to test the compatibility of the present $\omega_{eff}$ model with the $\Lambda$CDM model. For the $\Lambda$CDM model, the $Om(z)$ diagnostic should give a value of $\Omega_{m0}$.

\begin{figure}[h]
\centerline{\includegraphics[scale=0.68]{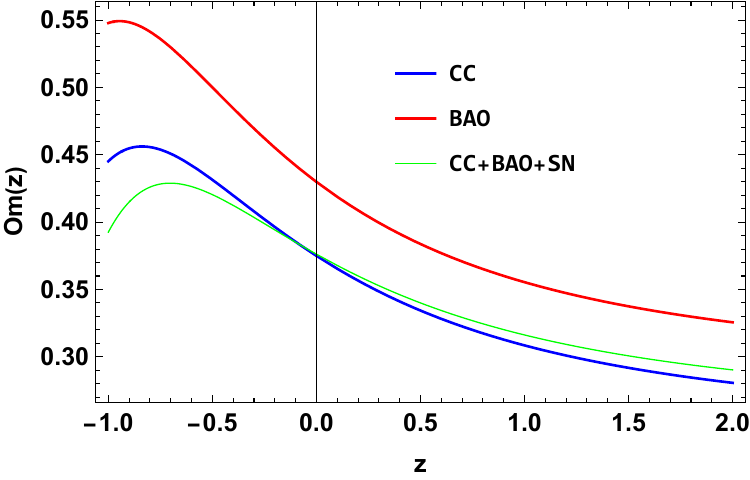}}
\caption{The plot displays the evolution of the $Om(z)$ diagnostic with respect to the redshift $z$ using the values constrained from the CC, BAO, and CC+BAO+SN datasets.}
\label{F_Om}
\end{figure}

For the present $\omega_{eff}$ model, Fig. \ref{F_Om} displays the behavior of $Om(z)$ for the values that are constrained from the CC, BAO, and CC+BAO+SN datasets. The behavior of the $Om(z)$ diagnostic is consistent with the behavior of the EoS parameter. Initially, the slope of $Om(z)$ is negative, indicating that the model behaves like a quintessence model. However, over time, the slope becomes positive, indicating that the model approaches the phantom region. Therefore, we can conclude that the model initially behaves like a quintessence model and eventually transitions to a phantom-type behavior.

\section{Conclusion}
\label{sec5}
The proposed model in this study is an attempt to understand the nature of DE, which is responsible for the current accelerated expansion of the Universe. To this end, the effective EoS parameter $\omega_{eff}$, which characterizes the properties of DE, has been parameterized in a specific way. The reason for choosing this particular functional form i.e. Eq. (\ref{eff}) is that it can effectively describe the two distinct phases of the Universe's evolution. At high redshifts i.e. $z\to\infty$, the Universe is dominated by matter and the effective EoS is $\omega_{eff}=0$. At low redshifts, i.e., at the present time $z=0$ and in the future as $z\to-1$, the Universe is dominated by DE and the effective EoS takes the values of $\omega_{eff}=-\frac{\alpha }{\alpha +\beta }$ and $\omega_{eff}=-1$, respectively. On the other hand, it is also possible to consider the EoS for DE from the perspective of non-perfect fluids, thus expanding the scope beyond the assumption of a perfect fluid \cite{Cardone}.

The present $\omega_{eff}$ model has then been tested against a variety of observational datasets, including the CC, BAO, and SN data. The purpose of the data analysis is to determine the constraints on the model parameters i.e. $H_{0}$, $\alpha$, and $\beta$ that are consistent with the observations. Specifically, the likelihood analysis has been performed using the MCMC method to obtain the posterior probability distribution of the model parameters. The constraints on the model parameters have then been derived from the posterior probability distributions of the parameters. Tab. \ref{tab1} summarizes the statistical analysis outcomes of the present $\omega_{eff}$ model. The table presents the values of the model parameters and their $1-\sigma$ and $2-\sigma$ CL, derived using different cosmological datasets such as the CC, BAO, and CC+BAO+SN.

The Hubble constant represents a vital cosmological parameter that defines the current rate of cosmic expansion. Our investigation resulted in the values of $H_{0}=67.84_{-0.66}^{+0.66}$, $H_{0}=67.84_{-0.72}^{+0.73}$, $H_{0}=67.70_{-0.63}^{+0.62}$ for CC,
BAO, and CC+BAO+SN datasets, respectively, which agree with the most recent observations from diverse cosmological probes \cite{Planck2020,Chen1,Chen2,Aubourg}. In addition, the study has also placed constraints on the EoS parameter for DE, which plays a crucial role in explaining the evolution of the Universe. The analysis has yielded a constrained values for $\omega_{DE}$ of $\omega_{0}=-0.91_{-0.02}^{+0.01}$, $\omega_{0}=-0.83_{-0.02}^{+0.02}$, and $\omega_{0}=-0.91_{-0.03}^{+0.03}$, indicating the quintessence behavior of the Universe \cite{Hernandez,Zhang}. Hazra et al. \cite{Hazra} analyzed various parameterizations of DE using recent observational datasets. The resulting EoS parameter values are $\omega_{0}=-1.005_{-0.15}^{+0.17}$ for CPL parametrization \cite{CPL1,CPL2}, $\omega_{0}=-1.14_{-0.09}^{+0.08}$ for SS (Scherrer and Sen) parametrization \cite{Scherrer}, and $\omega_{0}=-0.95_{non-phantom}^{+0.007}$ for generalized CG model \cite{Sen}. The agreement between the generalized CG model and the present $\omega_{eff}$ model is apparent, indicating their compatibility in describing the behavior of DE. However, the CPL and SS parametrizations seem to prefer a lower value for the EoS of DE. Also, these results are consistent with the outcomes of Capozziello et al. \cite{Capozziello22}, who employed a MCMC method to conduct a comprehensive cosmographic analysis utilizing data from BAO, Type Ia SN, and Gamma-Ray Bursts. Their study focused on a model-independent approach to understanding the dynamics of the universe, parameterized by the CPL model.

The evolution of the density parameter for matter and DE has been presented in Figs. \ref{F_Omega1}, \ref{F_Omega2}, and \ref{F_Omega3}, for different datasets. The figures indicate that as the Universe expands, the matter density parameter gradually decreases while the DE density parameter becomes dominant, leading to the acceleration of the Universe's expansion. Also, Fig. \ref{F_q} displays the deceleration parameter $q(z)$, which demonstrates that the present $\omega_{eff}$ model successfully reproduces both the early decelerated expansion phase and the current late-time accelerated expansion phase. The transition from decelerated to accelerated expansion occurs at a redshift between 0.5 and 0.8, consistent with the findings of \cite{Farooq0}. The present values of the deceleration parameter are $q_{0}=-0.44_{-0.03}^{+0.04}$, $q_{0}=-0.36_{-0.04}^{+0.04}$, $q_{0}=-0.44_{-0.04}^{+0.05}$ for the CC, BAO, and CC+BAO+SN datasets, respectively. For the same datasets, the values of the transition redshift are $z_{tr}=0.77_{-0.03}^{+0.02}$, $z_{tr}=0.61_{-0.01}^{+0.01}$, $z_{tr}=0.73_{-0.02}^{+0.01}$ \cite{Farooq,Jesus,Garza}.

Finally, we analyzed the geometrical parameters of the DE model. Figs. \ref{F_rq}, \ref{F_rs}, and \ref{F_Om} show the behavior of the state finder and $Om(z)$ diagnostics, which supported the quintessence-like behavior of DE in our model. The state finder diagnostics plot indicated $r<1$ and $s>0$, while the $Om(z)$ plot suggested a deviation from the $\Lambda$CDM model with a negative slope in the present. These findings provide further insight into the behavior of our model.

This approach of reconstruction of the effective or total EoS could also be extended to modified gravity models where the effective EoS may differ from that of the standard $\Lambda$CDM model. In particular, this could be used as future work to explore the behavior of modified gravity theories ($f(R)$ gravity, $f(T)$ gravity, and $f(Q)$ gravity) and distinguish them from DE models \cite{fR1,fR2,fT1,fT2,fQ1,fQ2,fQ3,fQ4,fQ5,fQ6,fQ7}. Moreover, such an approach could also provide insights into the nature of gravity itself, potentially leading to new theoretical developments in our understanding of gravity and the Universe as a whole.

\section*{Acknowledgments}
This work was supported and funded by the Deanship of Scientific 
Research at Imam Mohammad Ibn Saud Islamic University (IMSIU) 
(grant number IMSIU-RG23008).

\textbf{Data availability} This article does not introduce any new data.

\begin{widetext}
\section*{Appendix}
\label{app}

Density parameters:
\begin{equation*}
    \Omega _{m}\left( z\right) =\frac{\Omega_{m0} (z+1)^3 (\alpha +\beta )^{3/2} \exp \left(\frac{3 (\beta -1) \left(\tan ^{-1}\left(\frac{\beta +1}{\sqrt{4 \alpha -(\beta -1)^2}}\right)-\tan ^{-1}\left(\frac{\beta +2 z+1}{\sqrt{4 \alpha -(\beta -1)^2}}\right)\right)}{\sqrt{4 \alpha -(\beta -1)^2}}\right)}{(\alpha +\beta +z (\beta +z+1))^{3/2}},
\end{equation*}
\begin{multline*}
    \Omega _{DE}\left( z\right) =\left( \frac{3 H_{0}^2 (\alpha +\beta +z (\beta +z+1))^{3/2} \exp \left(-\frac{3 (\beta -1) \left(\tan ^{-1}\left(\frac{\beta +1}{\sqrt{4 \alpha -(\beta -1)^2}}\right)-\tan ^{-1}\left(\frac{\beta +2 z+1}{\sqrt{4 \alpha -(\beta -1)^2}}\right)\right)}{\sqrt{4 \alpha -(\beta -1)^2}}\right)}{(\alpha +\beta )^{3/2}}-\right. \\ 
     3H_{0}^2 \Omega_{m0} (z+1)^3 \Bigg) \left(\frac{(\alpha +\beta )^{3/2} \exp \left(\frac{3 (\beta -1) \left(\tan ^{-1}\left(\frac{\beta +1}{\sqrt{4 \alpha -(\beta -1)^2}}\right)-\tan ^{-1}\left(\frac{\beta +2 z+1}{\sqrt{4 \alpha -(\beta -1)^2}}\right)\right)}{\sqrt{4 \alpha -(\beta -1)^2}}\right)}{3 H_{0}^2 (\alpha +\beta +z (\beta +z+1))^{3/2}}\right).
\end{multline*}

DE EoS parameter:
\begin{multline*}
   \omega_{DE}(z) = -\Bigg(\alpha  \sqrt{\alpha +\beta +z (\beta +z+1)}\Bigg)\cr\left[\left(\Omega_{m0} z^3 (\alpha +\beta )^{3/2} \left(-\exp \left(\frac{3 (\beta -1) \left(\tan ^{-1}\left(\frac{\beta +1}{\sqrt{4 \alpha -(\beta -1)^2}}\right)-\tan ^{-1}\left(\frac{\beta +2 z+1}{\sqrt{4 \alpha -(\beta -1)^2}}\right)\right)}{\sqrt{4 \alpha -(\beta -1)^2}}\right)\right)\right)\right.\\
   + z^2 \left(\sqrt{\alpha +\beta +z (\beta +z+1)}-3 \Omega_{m0} (\alpha +\beta )^{3/2} \exp \left(\frac{3 (\beta -1) \left(\tan ^{-1}\left(\frac{\beta +1}{\sqrt{4 \alpha -(\beta -1)^2}}\right)-\tan ^{-1}\left(\frac{\beta +2 z+1}{\sqrt{4 \alpha -(\beta -1)^2}}\right)\right)}{\sqrt{4 \alpha -(\beta -1)^2}}\right)\right)\\   
   +(\alpha +\beta ) \left(\sqrt{\alpha +\beta +z (\beta +z+1)}-\Omega_{m0} \sqrt{\alpha +\beta } \exp \left(\frac{3 (\beta -1) \left(\tan ^{-1}\left(\frac{\beta +1}{\sqrt{4 \alpha -(\beta -1)^2}}\right)-\tan ^{-1}\left(\frac{\beta +2 z+1}{\sqrt{4 \alpha -(\beta -1)^2}}\right)\right)}{\sqrt{4 \alpha -(\beta -1)^2}}\right)\right)\\
   +z \left(\beta  \left(\sqrt{\alpha +\beta +z (\beta +z+1)}-3 \Omega_{m0} \sqrt{\alpha +\beta } \exp \left(\frac{3 (\beta -1) \left(\tan ^{-1}\left(\frac{\beta +1}{\sqrt{4 \alpha -(\beta -1)^2}}\right)-\tan ^{-1}\left(\frac{\beta +2 z+1}{\sqrt{4 \alpha -(\beta -1)^2}}\right)\right)}{\sqrt{4 \alpha -(\beta -1)^2}}\right)\right)\right. \\ \left.\left.
   -3 \alpha  \Omega_{m0} \sqrt{\alpha +\beta } \exp \left(\frac{3 (\beta -1) \left(\tan ^{-1}\left(\frac{\beta +1}{\sqrt{4 \alpha -(\beta -1)^2}}\right)-\tan ^{-1}\left(\frac{\beta +2 z+1}{\sqrt{4 \alpha -(\beta -1)^2}}\right)\right)}{\sqrt{4 \alpha -(\beta -1)^2}}\right)+\sqrt{\alpha +\beta +z (\beta +z+1)}\right)\right]^{-1}.
\end{multline*}

$Om(z)$ diagnostic: 
\begin{equation*}
    Om(z)=\frac{\frac{(\alpha +\beta +z (\beta +z+1))^{3/2} \exp \left(-\frac{3 (\beta -1) \left(\tan ^{-1}\left(\frac{\beta +1}{\sqrt{4 \alpha -(\beta -1)^2}}\right)-\tan ^{-1}\left(\frac{\beta +2 z+1}{\sqrt{4 \alpha -(\beta -1)^2}}\right)\right)}{\sqrt{4 \alpha -(\beta -1)^2}}\right)}{(\alpha +\beta )^{3/2}}-1}{(z+1)^3-1}.
\end{equation*}
\end{widetext}

\end{document}